\documentclass[iop]{emulateapj}
\usepackage{graphicx}
\slugcomment{}

\shorttitle{Stellar and ISM spirals}
\shortauthors{Wada et al.}

\begin{document}


%\title{\textcolor{red}{Interplay between  the ISM and self-excited stellar spirals in Galactic Disks} }
\title{{Interplay between Stellar Spirals and the ISM  in Galactic Disks} }


\author{Keiichi Wada}
\affil{Graduate School of Science and Engeering, Kagoshima University, Kagoshima, Japan}
\email{wada@cfca.jp}

\author{Junichi Baba}
\affil{Center for Computational Astrophysics, National Astronomical Observatory of Japan}
\email{baba.junichi@nao.ac.jp}

\and

\author{Takayuki R. Saitoh\altaffilmark{1}}
\affil{National Astronomical Observatory of Japan}

\email{saitoh.takayuki@nao.ac.jp}


\altaffiltext{1}{JSPS fellow}


\begin{abstract}
We propose a new dynamical picture of  galactic stellar and gas spirals, based on 
hydrodynamic simulations in a `live' stellar disk. 
{We focus especially on spiral structures excited in a isolated galactic disk without a stellar bar.}
Using
high-resolution, 3-dimensional $N$-body/SPH simulations, 
we found that the spiral features of the gas in galactic disks are formed by 
essentially different mechanisms from the galactic shock in stellar density waves.
The stellar spiral arms and the interstellar matter on average corotate in a galactic potential at
any radii. 
Unlike the stream motions in the galactic shock, the interstellar matter flows into the
local potential minima with irregular motions.  The flows converge to form dense gas clouds/filaments near the bottom of
the stellar spirals,  whose global structures resemble  dust-lanes seen in late-type spiral galaxies. 
The stellar arms are non-steady;  they are wound and stretched 
by the galactic shear, and thus local densities of 
the arm change on a time scale of $\sim 100$ Myrs, due to bifurcating or merging with other arms.
This makes the gas spirals associated with the stellar arms non-steady. 
The association of dense gas clouds are eventually dissolved into inter-arm regions with  non-cirucular motions.
Star clusters are formed from the cold, dense gases,  whose ages are less than $\sim 30$ Myrs, 
and they are roughly associated with the background stellar arms without
a clear spatial offset between gas spiral arms and distribution of young stars.
\end{abstract}



\section{Introduction}

{Spiral structures are the most prominent features in disk galaxies, which can be excited by tidal interactions with companion galaxies \citep[e.g.][]{Oh+2008, Dobbs+2010, Struck+2011}
as well as by the central stellar bars \citep[][(section 6.4)]{SellwoodSparke1988, BinneyTremaine2008}.
Spiral structures can also be self-excited and maintained without the gravitational perturbations 
in globally stable disks as proposed by \citet{LinShu1964}; They are interpreted as stationary density waves propagating in a stellar disk 
\citep[see also][]{BertinLin1996}.  In this paper, we focus on dynamics of the ISM in 
stellar spirals arisen in 
disk galaxies without external perturbations. }

In the conventional picture, the stellar density waves generate spiral perturbations in a galactic potential. 
Dynamics of the interstellar medium (ISM) in the spiral potential
have been theoretically studied since \citet{Fujimoto1968}, who found a standing shock,
called a `galactic shock',  in a tightly wrapped spiral potential.
Linear and non-linear behavior of the shock has been extensively studied from
various perspectives \citep[e.g.,][]{Roberts1969,Shu+1973, Woodward1975, Norman1978}.
The galactic spiral dust-lanes and associated star formation in the arms have since been regarded as a consequence of the galactic shock.
%Due to the radiative cooling, the ISM is strongly compressed in the shocked layer. 
{Although this is a widely accepted {\it static} picture of the galactic spiral, 
%{it is not clear whether the galactic shocks are dynamically stable.
%If they are unstable, a question arises: how is it consistent with the stationary picture?}.
%it is still unclear whether the stationary shocks are consistent with  formation processes of the molecular clouds near the spiral arms  \citep[e.g.,][]{Hartmann+2001,Vazquez-Semadeni2006, Heitsch+2009}, 
%which are in principle unstable phenomena and 
%necessary for subsequent star formation.
recent time-dependent, multi-dimensional simulations suggested that  shocked layers in spiral potentials are not always dynamically stable }
\citep[e.g.,][]{Kim+2006,ShettyOstriker2006,DobbsBonnell2006}\footnote{See also recent numerical simulations on tidally-driven spirals \citep{Oh+2008, Struck+2011}.}.  
 Spiral shocks can be hydrodynamically unstable (`wiggle instability' \citep{WadaKoda2004}), 
and sometimes form `spurs' \citep{KimOstriker2002}.
%if the Mach number and pitch angles are large enough.  
The substructures in the inter-arm regions  are 
formed as a result of formation of clumps in the layers.
%\footnote{Note that spurs are not `waves'. Clumps formed 
%by any kind of instabilities near the potential trough, such as gravitational instability, can be sheared off, and as a result elongated structures in the inter-arm regions are formed.}.  
%A similar phenomenon can be seen even in  `cloud-fluid' simulations by e.g. Tomisaka (1986). 
%The wiggle instability seen in the hydrodynamic simulations
%is essentially the Kelvin-Helmholtz instability, 
In reality, various physics, such as  self-gravity of the gas, radiative cooling, 
heating by stars, and the magnetic field  could affect dynamics and structures of the ISM 
in a spiral potential \citep[see][and references therein]{Kim+2010}. 
Using three-dimensional hydrodynamic simulations, taking into account 
the self-gravity of the gas and radiative cooling/heating processes and the supernova feedback, 
\citet{Wada2008} found that the classic galactic shocks are unstable and transient, 
and that they migrate to globally quasi-steady, 
inhomogeneous arms due to nonlinear development of gravitational, thermal, 
and hydrodynamical  instabilities. 

One unrealistic assumption in most previous 
hydrodynamic and magneto- hydrodynamic simulations of galactic spirals is, however, 
that the spiral potential is `fixed'.
Thus, dynamical interactions between the gas and stellar components were not considered\footnote{ \citet{DobbsBonnell2006} used a time-dependent gravitational potential taken from an $N$-body simulations of
\citet{SellwoodCarlberg1984}, and ran 3-D, two-phase (i.e., cold and warm gas only) Smoothed Particle Hydrodynamics  simulations. }. 
Yet,  $N$-body simulations have shown that 
self-induced spiral structures in stellar disks are not stationary \citep{SellwoodCarlberg1984,Sellwood2000,SellwoodBinney2002,Fuchs+2005,Fujii+2011, Sellwood2011}. 
This is also the case in an $N$-body disk with a central bar \citep{Baba+2009}.
In any $N$-body models, amplitudes of spiral arms change on 
a time scale of galactic rotation, or even shorter. Stationary 
spiral waves with small pitch angles have never
been numerically reproduced \citep{Sellwood2010, Sellwood2011}.
Each spiral arm in the $N$-body disk is transient, and recurrently reformed.  
\citet{Fujii+2011}  revealed a self-regulating mechanism that maintains 
multi-arm spiral features 
for at least 10 Gyrs in a pure stellar disk using three-dimensional $N$-body simulations.
They showed that the dominant spiral modes are time-dependent, reflecting a 
highly non-linear evolution of spiral density enhancements.
%Thereby spiral arms cannot be diminished within a cosmological time-scale provided that the
%number of particle is as large as  $10^6$ to avoid a numerical relaxation.

What we should explore next is dynamical interactions between 
the ISM and the live (i.e. time-dependent) stellar disk.
%In fact,  it has been pointed out that the ISM could play an important role in the density wave hypothesis.
%Due to interaction between the galactic shocks and the density waves, the exponential growth of 
%the waves is damped \citep{RobertsShu1972,Kalnajs1972}\footnote{\citet[See also section 6.4 in ][]{BinneyTremaine2008}.}.
Interactions between the ISM and the stellar component in disk galaxies 
have been extensively studied since the 1970s, and it was suggested that
the ISM is essential for dynamical structures in disk galaxies
\citep[e.g.,][]{RobertsShu1972,Kalnajs1972,JogSolomon1984,LinBertin1985}.  
However, fully non-linear evolution of the realistic multi-phase ISM 
in a live-stellar disk has not been investigated so far.

 In this paper, we study dynamics of the multi-phase, self-gravitating gas 
in a self-induced 
 stellar spiral arms in a galactic disk using Smoothed Particle Hydrodynamics (SPH) and
 $N$-body simulations. 
 For comparison, a model using a fixed spiral potential is also presented.
When radiative cooling and heating processes as well as the self-gravity of 
the gas are taken into account, it is essential to solve the whole three-dimensional
disk; otherwise global non-axisymmetric, inhomogeneous structures and their evolution are not
properly followed.
We also implement `star' formation from cold/dense gas and energy feedback
from supernovae, by which we can study what 
triggers formation of young stars in the ISM in spiral arms.

 One should note that the present simulations are not fully cosmological simulations,
 which are necessary to understand a full history of galaxy formation. 
Therefore, the numerical experiments here provide us with basic physical insight into 
 dynamics of spiral arms in isolated late-type spiral galaxies at the present universe.
Dynamics of spirals developed under an effect of the stellar bar will 
 appear elsewhere (Baba et al. in prep.).

%%%%%%%%%%%%%%%%%%%%%%%%%%%%%%%%%%%
\section{Numerical Methods and Model Setup}
\subsection{Methods}
\label{sec:method}

We used our original $N$-body/hydrodynamic simulation code {\tt ASURA}
\citep{Saitoh+2008,Saitoh+2009} to solve the Newtonian equation of
motions and the equations of hydrodynamics using the standard SPH methods \citep{Lucy1977,GingoldMonaghan1977,Springel2010}.
The numerical methods used here are the same as those in \citet{Baba+2009,Baba+2010}.
Here we briefly summarize them.

%\begin{eqnarray}
%\rho_i &=& 
%    \sum_j^{N_{\rm nb}}m_jW(|\mathbf{x}_i-\mathbf{x}_j|,h),\\
%\frac{d \mathbf{v}_i}{dt} &=& 
%     - \sum_j^{N_{\rm nb}}m_j
%     \left(\frac{p_i}{\rho_i^2} 
%             + \frac{p_j}{\rho_j^2} + \Pi_{ij}\right)
%     \nabla_iW(|\mathbf{x}_i-\mathbf{x}_j|,h)  \nonumber \\
%     && + \mathbf{g}_i - \nabla\Phi_{\rm DM}(\mathbf{x}_i),\\
%\frac{du_i}{dt} &=& 
%     \sum_j^{N_{\rm nb}}m_j
%     \left(\frac{p_i}{\rho_i^2}
%             + \frac{1}{2}\Pi_{ij}\right)
%     (\mathbf{v}_i-\mathbf{v}_j)
%             \cdot\nabla_iW(|\mathbf{x}_i-\mathbf{x}_j|,h)  \nonumber \\
%       &&  + \frac{\Gamma_i - \Lambda_i}{\rho_i},
%\end{eqnarray}
%where $m$, $\rho$, $p$, $u$, $\mathbf{v}$, $\mathbf{x}$, and $\Phi_{\rm DM}$
%are the mass, density, pressure, specific internal energy, velocity, 
%position of the gas, and the gravitational potential of the dark matter
%halo, respectively. 
%We assume an ideal gas EOS $p = (\gamma-1)\rho u$, 
%with $\gamma =5/3$.
%$W(x,h)$ and $h$ are the SPH smoothing kernel and the smoothing length, 
%respectively, and $h$ is allowed to vary both in space and time with
%the constraint that the typical number of neighbours for each particle
%is $N_{\rm nb} = 32\pm2$. 
%The artificial viscosity term $\Pi_{\rm ij} $ \citep{Monaghan1997} and
%the correction term to avoid large entropy generation in pure shear
%flows \citep{Balsara1995} are used.
%Radiative cooling of the gas  was solved assuming an
%optically thin cooling function with solar metallicity which covered a range of temperature, $20$ K through $10^8$ K
%\citep{WadaNorman2001}.  
%
The self-gravity of stars
and SPH particles  is calculated by the Tree method with GRAPE (GRavity PipE), 
a special purpose hardware for $N$-body simulations \citep{Sugimoto+1990}. 
Here we used a software emulator of GRAPE,
Phantom-GRAPE (Nitadori et al. in preparation).  
The gravitational softening length is $10~{\rm pc}$ for both SPH and $N$-body particles.

{We solve the energy equation implicitly 
with the cooling function \citep{SpaansNorman1997, WadaNorman2001}
assuming soloar metallicity, in which
various cooling processes\footnote{(1) recombination
of H, He, C, O, N, Si, and Fe; (2) collisional excitation of H$_{\rm I}$,
C$_{\rm I- IV}$, and O$_{\rm I- IV}$; (3) hydrogen and helium bremsstrahlung;
 (4) vibrational and rotational excitation of and H$_2$;   (5) atomic and molecular cooling due to fine-structure emission of C, C$^+$, and O and rotational line emission of CO.} are taken into account.  
%See also \citep{Saitoh+2008} for 
Heating due to photoelectric heating of grains and polycyclic aromatic hydrocarbons (PAH)
is considered, assuming
a uniform far-ultraviolet radiation (FUV) with the $G_0 = 1.0$ in the Habing unit
\citep{GerritsenIcke1997}.
 We also  implemented energy feedback from supernovae.
} 

{

The treatment of star formation and the heating due to
the SN feedback was the same as those in \citet{Saitoh+2008, Saitoh+2009}.
We adopted the simple stellar population (SSP) approximation,
with the Salpeter initial mass function \citep{Salpeter1955}
and the mass range of $ 0.1 - 100 M_\odot$ is assumed.     
The local star formation rate (SFR), $d \rho_*/dt$, is 
assumed to be proportional to the local gas density, $\rho_{\rm gas}$, 
and inversely proportional to the local dynamical time, $t_{\rm dyn} \sim 1/\sqrt{G \rho_{\rm gas}}$:
\begin{equation}
\frac{d \rho_*}{dt} = C_* \frac{\rho_{\rm gas}}{t_{\rm dyn}}, \label{eq:sf}
\end{equation}
where $C_{*}$ is the dimensionless star formation efficiency parameter. 
The value of this parameter is usually calibrated by the global star formation properties, 
the Schmidt-Kennicutt relation \citep{Kennicutt1998,MartinKennicutt2001}.
An SPH particle
was replaced with a star particle following the Schmidt
law (Schmidt 1959) with a local star formation efficiency
$C_{\ast}=0.033$ in a probabilistic manner \citep{Saitoh+2008,Saitoh+2009,RobertsonKravtsov2008},
if criteria (1) $n_{\rm H} > 100~{\rm cm^{-3}}$, (2) $T < 100~{\rm K}$, 
and (3) $\nabla\cdot \vec{v}<0$,  are
satisffied.
Note that global
(galactic) star formation rate is not directly proportional
to the local star formation effciency, $C_{\ast}$ , 
but is rather controlled by the global evolution of the ISM \citep{Saitoh+2008}.
%The second point is specially important for numerical simulations of galaxy formation
%and evolution since our result does not depend strongly
%on the value of  $C_{\ast}$ used. 

We implemented type-II SN
feedback, where the energy from SNe is injected to the
gas around the star particles in the thermal energy. Each
SN releases $10^{51}$ ergs of thermal energy for the surrounding SPH 
particles (typically $32\pm 2$) particles.  
One gas particle ($m \sim 4000 M_\odot$) produces about 30 SNe during 20 Myrs.
This energy injection produces expanding cavities of hot gas ($T \sim 10^{6-7}$ K)
in the ISM \citep[see also section 2 in][]{SaitohMakino2010}.
%and therefore the energy and momentum feedback from SNe 
%are reproduced.  
Although the dynamics of the ISM on a galactic scale is mainly
determined by the gravitational force,  local structures of the ISM on a sub-kpc scale
are affected by the feedback from SNe.

}

\subsection{Live Spiral Model}

We first evolve a pure $N$-body stellar disk with an exponential 
profile in a static dark matter (DM) halo potential whose 
density profile follows the Navarro-Frenk-White profile \citep{Navarro+1997}.  
The initial stellar disk is generated using the Hernquist's method \citep{Hernquist1993}, and
 then it is evolved for a few Gyrs, 
randomizing azimuthal positions of stellar particles every several time-steps
to prevent the growth of nonaxisymmetric features 
\citep{McMillanDehnen2007,Fujii+2011}.
Figure 1 shows the initial circular velocity, velocity dispersions, and Toomre's Q value as a function of the galactocentric distance. 
Once  the stellar disk becomes almost featureless, 
we added a gaseous component, whose total mass is 10\% of the stellar disk.
This is regarded as an `initial' condition in the result presented in \S 3.
The model parameters of the DM halo, stellar disk, and gas disk are
summarized in Table \ref{tbl:model}. 
%The circular velocity curve is shown in Figure \ref{fig:angfreq}a.

The total numbers of $N$-body and SPH particles are initially $3\times 10^6$ and
$10^6$, respectively.  They are finally $3.2\times 10^6$ and $0.8 \times 10^6$, respectively.
 The corresponding particle masses are $11000~M_\odot$ and 
$3200~M_\odot$, respectively, owing to the star formation.  \citet{Fujii+2011} showed using pure $N$-body simulations that one million $N$-body particles are necessary to reproduce long-lived ($> 10$ Gyrs) spirals, and the results
do not change in a model with $N = 3\times 10^7$ particles.
\citet{Saitoh+2008}  explored the effect of the numerical resolutions in {\tt ASURA}
in terms of representing inhomogeneous structures of the multi-phase gas on a galactic scale, 
and concluded that  complicated structures of the multi-phase gas 
can be resolved with a particle mass of 3500 $M_\odot$, or smaller.
It was also confirmed that the results are converged between models with  3500 $M_\odot$
and 350 $M_\odot$.
%
% is necessary to resolve 
%the gravitational fragmentation of gas with $\sim
%10^5~M_\odot$ or larger,  in comparison with a model using ten times finer resolutions.

%%%%%%%%%
\subsection{Rigid Spiral Model}

In order to study the effect of the live stellar disk, 
we run a test model, in which the stellar disk 
is replaced with a time-independent spiral potential.
The gravitational potential of the stellar disk is
\begin{eqnarray}
\Phi_{\rm sp}(R,\phi,z) &=& \Phi_{\rm disk,0}(R,z)
\epsilon_{\rm sp}\frac{z_{\rm 0}}{\sqrt{z^2+z_{\rm 0}^2}} \\ \nonumber
& \times & \cos \left[m \left(  \phi+\cot i_{\rm sp}\ln\frac{R}{R_{\rm 0}}\right) \right],
\end{eqnarray}
on a rotating frame of the spirals. 
Here $\Phi_{\rm disk,0}$, $m$, $\epsilon_{\rm sp}$, $i_{\rm sp}$, and $z_{\rm 0}$ 
are the axisymmetric potential produced by the stellar disk, 
the number of stellar spiral arms, the strength of the spiral perturbation,  the pitch angle, 
and the scale-height, respectively.
We set $i_{\rm sp}=30^{\circ}$, $m =4$,  and $\epsilon_{\rm sp} = 0.02$. The phase is $R_0 = 1$ kpc and 
the pattern speed, $\Omega_{\rm sp}$, is $\simeq 12.2~{\rm km~s^{-1}kpc^{-1}}$
 ($R_{\rm CR} = 14$ kpc).
These parameters are chosen from typical structures and 
dynamics of the live spiral model.

%%%%%%%%%

%% Model Parameters
\begin{table}[htdp]
\caption{Model parameters for each mass component (dark matter halo, stellar
        disk, and gas disk). $c_{\rm NFW}$ is a concentration parameter in the Navarro-Frenk-White profile.}
\begin{center}
\begin{tabular}{ lll }
\hline
\hline 
 Component	&	Parameters	& Value	\\
\hline
Dark Matter Halo			& Mass	&	$6.3\times 10^{11}$ M$_{\odot}$\\
								& Radius 	&  $122$ kpc\\
								& $c_{\rm NFW}$ 	& $5.0$ \\
\hline
Initial Stellar Disk	& Mass	&  $3.2\times 10^{10}$ M$_{\odot}$\\
								& Scale Length 	& $4.3$ kpc \\
								& Scale Height	& $0.3$ kpc\\
\hline
Initial Gas Disk	& Mass	& $3.2\times 10^{9}$ M$_{\odot}$\\
								& Scale Length	& $8.6$ kpc\\
								& Scale Height	& $0.1$ kpc \\
\hline
\end{tabular}
\end{center}
\label{tbl:model}
\end{table}%

%%%%% results %%%%%%%%%%%
%
\section{Results}
%
%%%%%%%%%%%%%%%%%%%%
\subsection{Non-steady Spiral Arms and Its Kinematics}
Figure \ref{fig1} shows four snapshots of $V$-band surface brightness 
reproduced by the stellar density during 975 Myrs.   Obviously spiral arms are not stationary
on a time scale of  a few 100 Myrs (see also the movie in {\it the supplementary data}).
If we track one of the spiral arms, 
we observe that density of the arm is no longer constant --
 a part of the arm becomes faint or denser within one rotational period.
There are no steady grand-design spiral arms, 
in fact spiral arms  are  `wound up'.
Any arms are stretched by the galactic shear and attenuated.
Splitting, bifurcating, and merging of arms are often observed.
At the same time, new spiral arms are generated somewhere in the disk
and as a result there are always several spiral arms in the disk.
These phenomena suggest that the spiral arms do not propagate in the stellar disk
 as stationary density waves. 
 In the present model, the number of spiral arms is 4-6, but the dominant modes are
time- and radially-dependent
\citep[see also][for dynamical evolution of pure stellar spiral arms]{Fujii+2011}.

In the left panel of  Figure \ref{fig2}, 
the rotational frequency of the most dominant mode of the spirals is shown
as a function of radius at $t= 1.625 $ Gyrs.
It clearly shows that spiral arms mostly follow the galactic rotation $\Omega(R)$ at any radii.
They do not rotate with a constant pattern speed.
This is always the case during the evolution, as shown in the right three panels, in which time evolutions of  stellar densities  at three different radii, i.e. $R =$ 4.3, 8.6, and 12.9 kpc, are plotted.
The inclination of black/gray stripes represents rotational speed of the spiral arms,
showing that they basically follow the galactic rotation (red dashed line).
Each stripe changes its shading in several 100 Myrs, indicating local density of 
the arm does change.
 The radially dependent rotational speed of spiral arms 
 seems to be consistent with recent analysis of stellar kinematics in 
M51 and NGC 1068, in which 
a radially declining `pattern' speed of spiral arms
is found using the Tremaine-Weinberg method 
(Merrifield et al. 2006; Meidt et al. 2008).
The numerical result clearly shows that spiral arms do not 
propagate as density waves with a single pattern speed in the disk, but 
they rather behave like `material' arms\footnote{Since the arms consist of stars, 
and the stars are not confined in each arm for many rotational periods, 
the spiral-like density enhancements also possess some similar properties in the 
ideal density waves.
Detailed analysis of stellar orbits and their relationship to the non-steady spiral arms will be
discussed elsewhere (Baba et al. in prep.).}.

\subsection{Behavior of the ISM in the Non-steady Spiral Potential}

In these non-stationary stellar spiral arms, the structure and dynamics of the multi-phase ISM are
quite different from the prevalent  picture of the galactic shock.
Figure \ref{fig4} is a snapshot of cross-sections of the disk at three different galactic radii, 
revealing a typical azimuth distribution of  stellar and densities.
Relatively smooth curves represent azimuth distributions of 
the stellar density; for example there are 
four arms at $R= 8.6$ kpc and 6 $\sim$ 7 arms at $R = 12.9$ kpc
at $t = 1.441$ Gyrs.
{Total gas density}, on the other hand, is more spiky.
Each peak corresponds to a high density clouds or internal substructures of them.
It is clear that high density peaks of the gas tend to be associated with
 stellar arms.
Moreover, this complicated structure of 
the gas is not steady  (see also the movie in {\it the supplementary data}).
These features are essentially different from 
 the standing galactic shocks in a non-selfgravitating, isothermal gas in
the spiral potentials.

The concentrations of cold ($T < 100 $ K) gas seen in Fig. \ref{fig4} actually 
form spiral structures in a two-dimensional distribution.
Figure \ref{fig3} shows distribution of the cold gas ($T < 100 $ K) overlaid with the stellar density.
It is clear that gas clouds roughly
trace high density regions of stars, and they form complicated substructures
in the stellar spirals.  The associations of dense clumps 
can be traced as a single arm connected intermittently from the
central region to the outer disk.
As clearly seen in the movie in the supplementary data, 
a grand-design spiral arm of the gas is temporalily formed, but
it becomes sparse in a few $\sim 100$ Myrs.

As shown above, if the multi-arm stellar spirals are developed 
are developed by the swing amplification of density disturbances
in the galactic disk \citep{Toomre1981, Fujii+2011}, 
both the gas and stellar arms follow the local galactic rotational velocity on average, and as a result
the velocity of the gas relative to the stellar arms should be  $\sim 15$ km s$^{-1}$ or smaller.
On the rotating frame of the spiral arms, 
the ISM falls into the spiral potential from
both sides, that is there are flows of cold gas slower and faster than the stellar arm.
The flows converge into 
condensations of cold gas near the bottom of the potential well.
The relative velocities of the gas to the stellar spirals are comparable to the
random motion of the gas. In other words, the turbulent ISM  `sub-sonically' move near the
spiral potential\footnote{ {Effective sound speed, e.g. $\sqrt{c_s^2 + \sigma_v^2}$, where $\sigma_v$ is
velocity dispersion of the gas, can be larger than the thermal sound speed $c_s$ especially for cold gas.
Therefore the clumpy gas flow can be regarded as} {\it sub-sonic} relative to the
spiral arms, {in other words, 
effective pressure of the turbulent gas could dominate dynamics of the flow near the spiral arms,
even if the local Mach number is larger than unity (see discussion below).}}.
 This is an essential difference in the flow in a time-independent spiral potential.

{The converging gas flows near the spiral arms are  clearly seen 
in a two-dimensional map}, which is distinct from those in 
a rigid spiral potential.
Figure \ref{fig6} compares the velocity field of the cold gas ($T < 100$ K) between the live and rigid spiral potential 
models.
In the live spiral model, the velocity vectors of the gas 
on a galactic rotation frame are irregular, and they tend to
converge into the dense gas arms.  
In other words, since the relative velocity of the 
gas to the stellar spiral is small, 
 the gas is `trapped' by  the potential troughs,
forming high density regions.
This is in contrast to what is observed in the rigid spiral model, where 
the potential is given as described in \S 2.2.
The gas flows in the rigid spiral model are relatively regular, and
they change their directions near the gas arms, as typically seen in
 the two-dimensional galactic shock \citep{WadaKoda2004}.
 In both the models,  supernova feedback is included, but this is not
 essential to change the velocity field of the gas near the spiral potential, 
{although it sometimes dissolves the association of gas clouds (see \S\ref{sec3.3})}.
One should note here that the gas velocities relative to the
spiral potential (here we assume that the live stellar arms
follow the galactic rotation) are much {slower} than
those in the rigid spiral model. 
In the right panel of Fig. \ref{fig6}, 
frequency distributions of the Mach number $\cal{M}$ $ \equiv |v|/c_s$ in both models are shown,
where $c_s$ is sound velocity, $c_s \equiv (\rho_g k T_g/\mu)^{1/2}$ with the gas density $\rho_g$ and temperature $T_g$  of each SPH particle, 
mean molecular mass $\mu$, and $|v|$ is the gas local velocity on a galactic rotation (live spiral model)
or on a rotating frame of the spiral potential (rigid spiral model).
{The sound velocity ranges from $\sim 0.5$ to $\sim$ 300 km s$^{-1}$ depending on the local gas temperature. Therefore $\cal{M}$ represents Mach number of the ISM relative to the spiral potential.}
It is clear that the gas flows in the rigid spiral potential are highly supersonic (i.e. $\cal{M}$ $ \sim 10-40$).
On the other hand, in the live spiral model, the median is $\cal{M}$ $ \sim 2$.
{ In Fig. \ref{fig7new}, we plot the velocity field of the cold gas {\it relative to the 
galactic rotation}, i.e. mean background flow,  in the rigid model.  
Although the median is $\cal{M}$ $ \sim 3$ and there are more gas with higher Mach number ($\cal{M}$ $> 10$), 
the entire distributions of $\cal{M}$ are similar in both model.
Therefore it is reasonable that  the local density field of the gas in both models are similar, showing filamentary and clumpy 
morphology. }

If the converging gas flows of clumpy gas collide  with low Mach number near the spiral arms,  
it could cause weak shocks {\it locally.}
Weak shocks in warm gases near the stellar arms can trigger formation of molecular clouds 
through thermal instabilities \citep{InoueInutsuka2009, Vazquez-Semadeni2011}.

{ Each stellar arm becomes faint or
merges into other arms on a time scale of the galactic rotation, or even shorter.  
Therefore one important aspect of the gas in the non-steady stellar spirals is that 
the gas concentrations, like giant molecular associations (GMA),  formed in the 
 potential troughs can obtain additional kinetic energy due to
the variation of  the background stellar spiral potential. This can be a source of dissolving 
the gas arms and generating kpc-scale non-circular motions of the ISM (Figure \ref{fig6}).
}
We found that most of the
 cold gas clouds have  large `epicyclic' motions of which
 diameter is  comparable to typical intervals between spirals arms, i.e. $\sim 2-3$ kpc.
  In other words, 
 the gas circulates between spiral arms (see the movie in the supplementary data). 
  
%  These flows of cold gas often converge near the stellar potential, by which
 % GMA-like structures are formed by collisions of cold streams.
  The large peculiar motions of the gas are in fact observed as  
proper motions of maser sources in our Galaxy \citep{Reid+2009b}. 
The observed peculiar motions of the star forming regions are 
quantitatively consistent with a numerical barred-spiral galaxy
constructed by the same methods as those in the present paper \citep{Baba+2009}.
In Figure \ref{fig7}, an example of such a collision between cold streams is shown.
The sequence of snapshots during  80 Myrs shows that
concentrations of cold gas clumps,  each of which is initially $\sim $ 1 kpc away from the other,
merge into one association of dense gas near the center of
the stellar arm.
The massive concentrations of cold gas, which 
 would correspond to GMA in 
 real galaxies, are temporarily formed and eventually dispersed as
 the background stellar arms become faint.

\subsection{Star Formation and its Feedback in Spiral Arms}
\label{sec3.3}

Here we implemented `star formation' from cold, dense gas.
Thus, collisions of dense gas clouds shown in Figure \ref{fig6} may trigger star formation.
Once the star particles are generated from their mother clouds based on the star-formation 
criteria described in \S 2, 
 they behave as additional collisionless $N$-body 
particles (we call them `young stars') in the live stellar disk.
Figure \ref{fig8} shows entire structures of old (i.e. disk stars) and young stars of which
ages are less than 30 Myrs as well as the 
cold gas at {$t = 1.6$ Gyrs}.
The young stars tend to form clusters along with
the spiral arms consisting of old stars.  However,  
there is no clear offset between young stars and the stellar spirals (i.e. gravitational potential), 
in contrast to the prediction of the galactic shock hypothesis. 
This is reasonable from the distribution and kinematics of cold, dense gas shown in Figs \ref{fig4} and \ref{fig3}.

{ In Figures.  \ref{fig9a}  and \ref{fig9b}, evolution of a spiral arm for 
a short period ($20$ Myrs) is shown by density and temperature maps, respectively.
The gas clouds in the white circle around $R = 7.5$ kpc are cold ($T_g < 100 $ K), and are surrounded by
warm ($T_g\sim  10^3 - 10^5$ K), less dense gas.  Around the cold clouds,  several hot spots
($T_g \sim 10^6$ K) caused by supernovae appear, and 
the dense clouds are partially  dissolved.  Figure \ref{fig11} shows  a
frequency distribution of `life time' of high density gas, for randomly selected 1000 SPH particles.
It shows that most gas particles do not keep  $n_{\rm H} > 100$ cm$^{-3}$ 
for more than 10-20 Myrs.
This reflects the fact that the high density clumps produce massive stars, and 
they are eventually destroyed by SNe originated in the clumps.
Typically a 4000 $M_\odot$ SPH particle produces about 30 SNe during 30 Myrs.
As seen in Figure \ref{fig8},  gas filaments consisted of dense gas complexes 
in the stellar spiral arms produces clusters of massive stars, 
and energy feedback from SNe originated in
the clusters sometimes trigger expanding motions of the gas clouds (see also the movies in 
the supplementary data).  
 }

\subsection{Spurs}

{Many spiral galaxies, especially the ones with well defined primary dust-lanes, exhibit
so called spurs and feathers \citep{Elmegreen1980, Scoville+2001,LaVigne+2006}
On the other hand, 
although the gas spiral arms shows rich substructures, 
we do not see clear periodic spurs in the inter-arm regions in the present model (Fig.\ref{fig8}).
%The gas flows on the galactic rotating frame are more irregular than those in the rigid model (see Fig. \ref{fig6}). 
This is in contrast to results in previous hydrodynamic and magneto-hydrodynamic simulations
where isothermal gas is evolved in a fixed spiral potential with a constant pattern speed, $\Omega_p$
\citep{KimOstriker2002, WadaKoda2004, ShettyOstriker2006, DobbsBonnell2006, KimOstriker2006}. 
This implies that difference of rotational velocities between spiral potentials and the ISM
is essential for forming the downstream spurs/feathers.
In the present model,  there are no strong shears or ordered motions of the gas in the live spiral model, 
because both the stellar spirals and the ISM follow the galactic rotation, i.e. $\Omega_p \sim \Omega(R)$.
If $\Omega_p  < \Omega(R)$, as suggested by recent numerical simulations of 
tidally-excited spirals \citep{Dobbs+2010, Struck+2011}, spurs could be formed.
This is consistent with that spurs tend to be clearly observed in well-defined two-arm spirals, such as 
M51 \citep{Scoville+2001}. 
%Finally one should note that although \citet{LaVigne+2006} found 
%periodic spurs/feathers in about 20\% of  images of spiral galaxies taken by the Hubble Space Telescope,
%it is hard to distinguish between feathers/spurs originated in the instability of galactic shocks and inter-arm filaments. The cold ISM intrinsically have quasi-periodic filamentary
%structures in galactic disks \citep{WadaNorman2001}. 
}

\section{Conclusions}
We propose a new picture of  galactic stellar and gas spirals, based on 
hydrodynamic simulations in a `live' stellar disk.  
{Here we focus on basic properties of multi-arm spiral structures in isolated disk galaxies
without a central stellar bar.  Tidally excited spirals in interacting galaxies are 
probably produced by the kinematic density wave \citep{Kalnajs1973} beyond
scope of this paper \citep[see e.g.][]{Oh+2008, Dobbs+2010, Struck+2011}}.¡¡In contrast to the theory of standing galactic shocks \citep[e.g.][]{Fujimoto1968,Roberts1969, Shu+1972}
in static spiral potentials caused by stellar
density waves  \citep{LinShu1964,BertinLin1996}, 
high-resolution, three-dimensional $N$-body/SPH simulations using the {\tt ASURA} code \citep{Saitoh+2008}
 revealed non-steady and non-linear evolution of gas and stellar spirals.
Morphology of the spiral structures looks similar to that seen in late-type spiral galaxies.
Important features found in the numerical simulations and their implications 
are summarized as follows:

\begin{enumerate}
\item  Rotational speeds of stellar spiral arms follow the galactic rotation. 
Grand-design spirals therefore cannot maintain their shapes over a rotational period.
As a result they are wound up and  torn off by 
the galactic shear.   Spiral arms are recurrently formed, and often merge into other arms, by which
local densities of the arm are no longer constant.
This non-steady nature of spiral arms has been reported  in 2-D and 3-D $N$-body simulations of pure stellar disks 
\citep[e.g.,][]{SellwoodCarlberg1984, Sellwood2011, Fujii+2011}. We confirmed that this is also the case in a
stellar disk with the ISM.
\item The gas component on average also follows the galactic rotation. 
As a consequence, the Mach number of the gas relative to {spiral potentials}
is around $\sim 2$,  while on the other hand, it is more than 10 in the rigid spiral model.
Flows of cold, dense gas converge toward the stellar arms, resulting in formation of massive associations of the cold gas and star formation.
This is an essential difference from the classical galactic shock, where the gas passes through the potential and 
forms a oblique, global shock along the spiral potential.
\item {Most of the cold and dense gas clouds are
 associated with the time-dependent stellar spirals.  
 The morphology of the gas spirals look similar to 
 observed dust-lanes in multi-arm spiral galaxies.
We, therefore, suspect that dust-lanes in late-type spiral galaxies,
especially those in multi-arm spirals, are not galactic shocks.}

\item {Dense gas regions in the bottom of stellar arms are eventually dissolved into inter-arm regions.
They are affected by
variation of  the background stellar spiral potential on a time scale of $\sim 100$ Myrs. 
This can be a source of dissolving 
the gas arms and generating non-circular 
motion of the ISM.  Energy feedback from supernovae also contributes to dissolve the 
complex of dense gas clumps.
The entire motions of the cold gas on a galactic rotating frame are `epicyclic', but whose radii are comparable to
intervals between spiral arms.}

\item The mechanism presented here may also explain 
complicated sub-structures of dust-lanes and distribution of molecular gas 
in spiral galaxies, such as M33 \citep[][]{Onodera+2010,Gratier+2010}, or
our Galaxy. 
In fact, \citet{Baba+2010} showed using the same method that features
  in observed position-velocity diagrams of CO and HI in our Galaxy
 are identified as clumpy gas spiral arms. 
 
\item In the model, stars are formed from cold, dense gases. Consequently,  
young stars, whose ages are less than $30$ Myrs, form clusters, and 
they are associated with the background stellar arms.
We do not see clear spatial offset between gas spiral arms and distribution of young stars.
%\item GMAs are formed through collisions between circulated flows near the spiral potential.
%\item Gas is NOT necessary to keep the long-lived stellar spirals. It could temporarily enhance them.
%\item The dynamics of the ISM and stellar spiralsare far from 
\end{enumerate}

The results above imply that it is essential to solve both live stellar disk and the multi-phase ISM 
self-consistently in order to understand dynamical structures in spiral galaxies 
and their evolution. 
%Although some of them were already recognized by theorists since 1960s, 
%recent progress in numerical techniques more clearly showed the dynamical nature of
%the galactic spirals
%in a more self-consistent manner.  
If we change parameters of model galaxies, such as the rotation curve and 
disk mass, we can reproduce 
spiral galaxies with different morphology. It is also important to realize that
structures in spiral galaxies should be `time-dependent'.  Some differences in morphologies of isolated spiral galaxies, 
such as the number of spiral arms and substructures in spirals could be explained 
by evolutional features.
Barred spirals, in which $m=2$ arms are driven by
a stellar bar,  will be separately discussed elsewhere (Baba et al., in prep).

\acknowledgments

The authors are grateful to J. Makino and E. Kokubo for stimulating discussions. 
We are also grateful to the referee for valuable comments. 
Numerical simulations 
were performed by Cray XT-4 in the Center for Computational Astrophysics (CfCA), 
National Astronomical Observatory of Japan. 
T.R.S. is financially supported by a Research Fellowship from the Japan Society for the Promotion of Science for Young Scientists.

%%%%%
% Figures
%%%%%
%%% Figure 1 %%%%%%
\begin{figure*}
\begin{center}
 \includegraphics[width=0.90\textwidth]{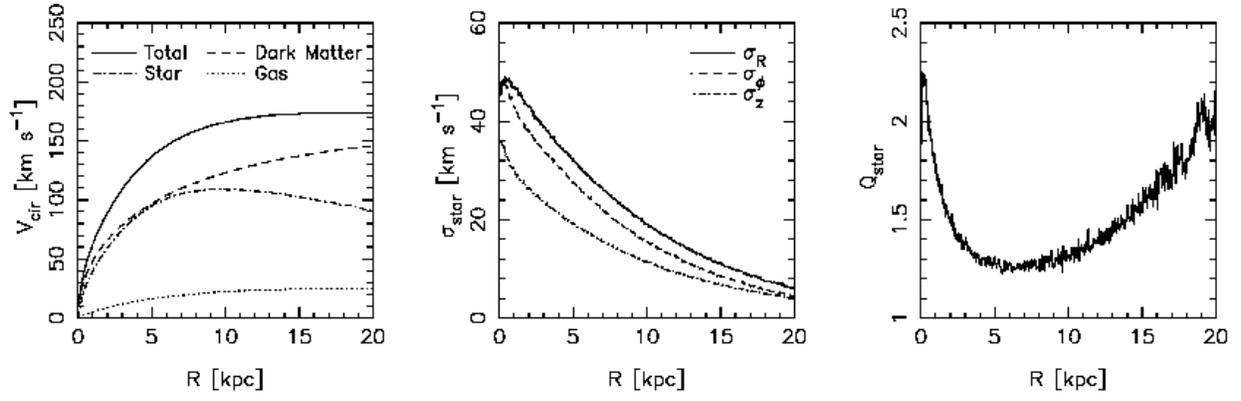} 
 \caption{Initial condition of the stellar disk models in the live spiral model.
(left) The circular velocity and contributions of individual component as a function of the galactocentric distance, $R$. (center) The velocity dispersions in the cylindrical coordinates ($R,\phi,z$) as a function of $R$. (right) Toomre's $Q$ value as a function of $R$.
 }
 \label{fig:InitialCondition}
\end{center}
\end{figure*}

%%% Figure 2 %%%%%%
\begin{figure}[t]
% \vspace*{-2.0 cm}
\begin{center}
 \includegraphics[width=0.35\textwidth]{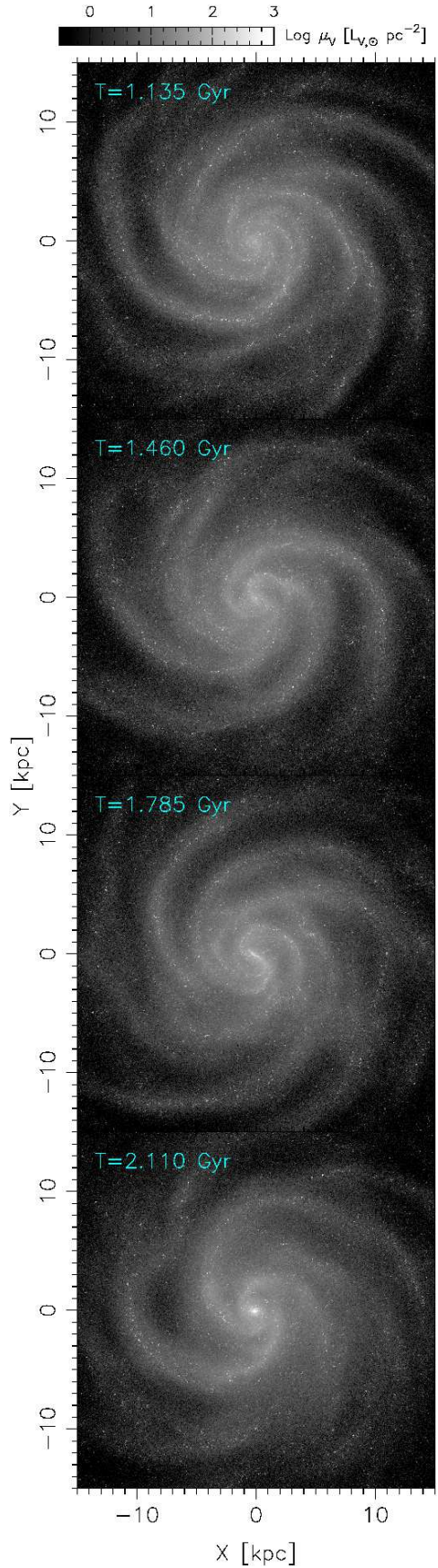} 
% \vspace*{-1.0 cm}
 \caption{
% 	Evolution of stellar spirals on a rotation frame of $R=2R_{\rm sd}$, here $R_{\rm sd}$ is the scale length of the 
%	initial stellar disk (table 1). 
	Evolution of stellar spirals on a rotation frame at $R=8.6$ kpc.	Gray-scale represents  $V$-band surface luminosity calculated from stellar density with
 a population synthesis model, PEGASE.2 \citep{FiocRocca1999}, assuming a solar metallicity
 and the  Salpeter initial mass function. }
 \label{fig1}
\end{center}
\end{figure}

%%% Figure 3 %%%%%%
\begin{figure*}[t]
% \vspace*{-2.0 cm}
\begin{center}
\includegraphics[width=1.0\textwidth]{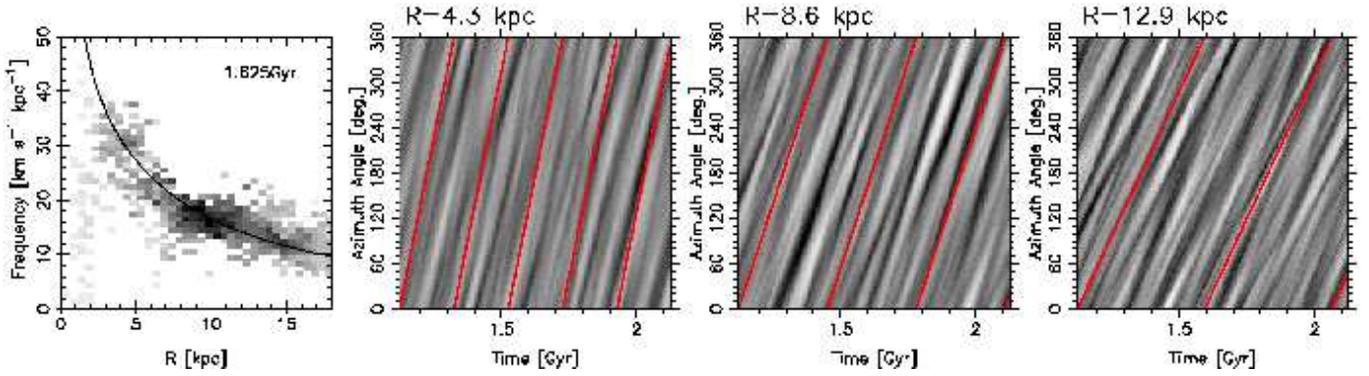} 
% \vspace*{-1.0 cm}
 \caption{
 (left) Rotational frequency of the most dominant mode {($m=4$)} 
 of the spirals is shown as a function radius at a snapshot of $t= 1.625 $ Gyr. 
{The gray scale represents the relative amplitude of the mode}.
 {The solid curve represents   the circular  frequency $\Omega(R)$}.  (right three panels)
 Kinematics of stellar arms at $R= 4.3, 8.6$ and 12.9 kpc.  
 Black/gray stripes represent how the azimuthal positions of high density regions of stars move. 
 The red solid lines represent the galactic rotation at  each radius.}
   \label{fig2}
\end{center}
\end{figure*}

%%% Figure 4 %%%%%%
\begin{figure*}[t]
% \vspace*{-2.0 cm}
\begin{center}
  \includegraphics[width=0.90\textwidth]{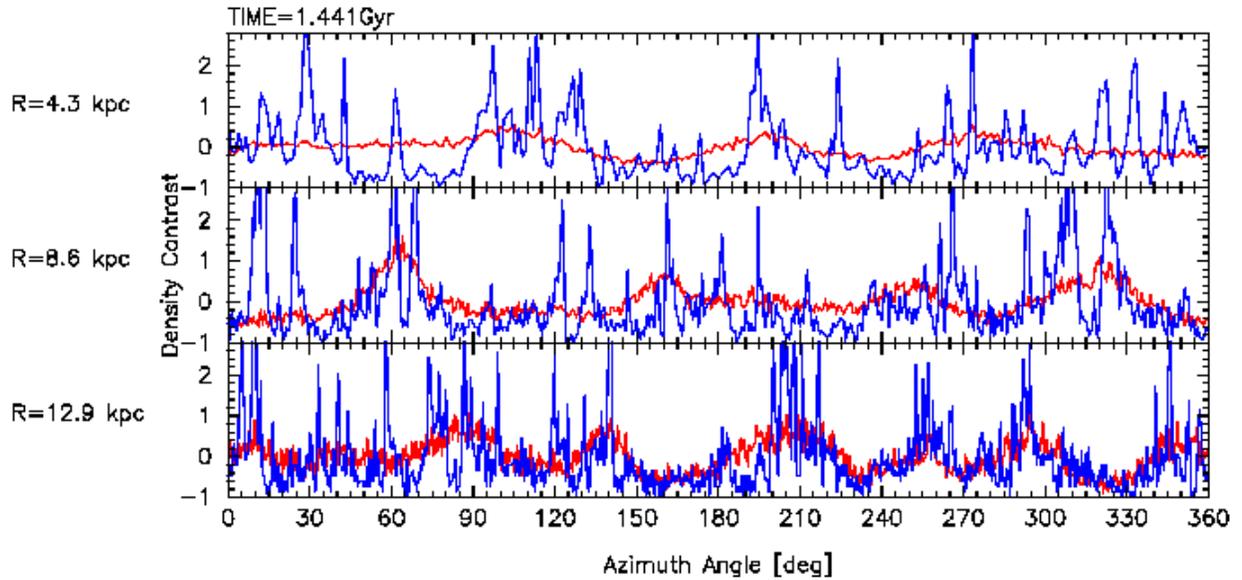} 
% \vspace*{-1.0 cm}
 \caption{Azimuth distributions of stellar (red) and total gas (blue) density contrasts at $R= 4.3, 8.6$ and $12.9$ kpc
 at $t= 1.441$ Gyr.
 See also the movie in {\it the supplementary data} for time evolution, where
 azimuthal density distributions are plotted on a rotating frame of a galactic rotation at 
 each radius.  }
   \label{fig4}
\end{center}
\end{figure*}

%%% Figure 5 %%%%%%
\begin{figure}[t]
\begin{center}
\includegraphics[width=0.50\textwidth]{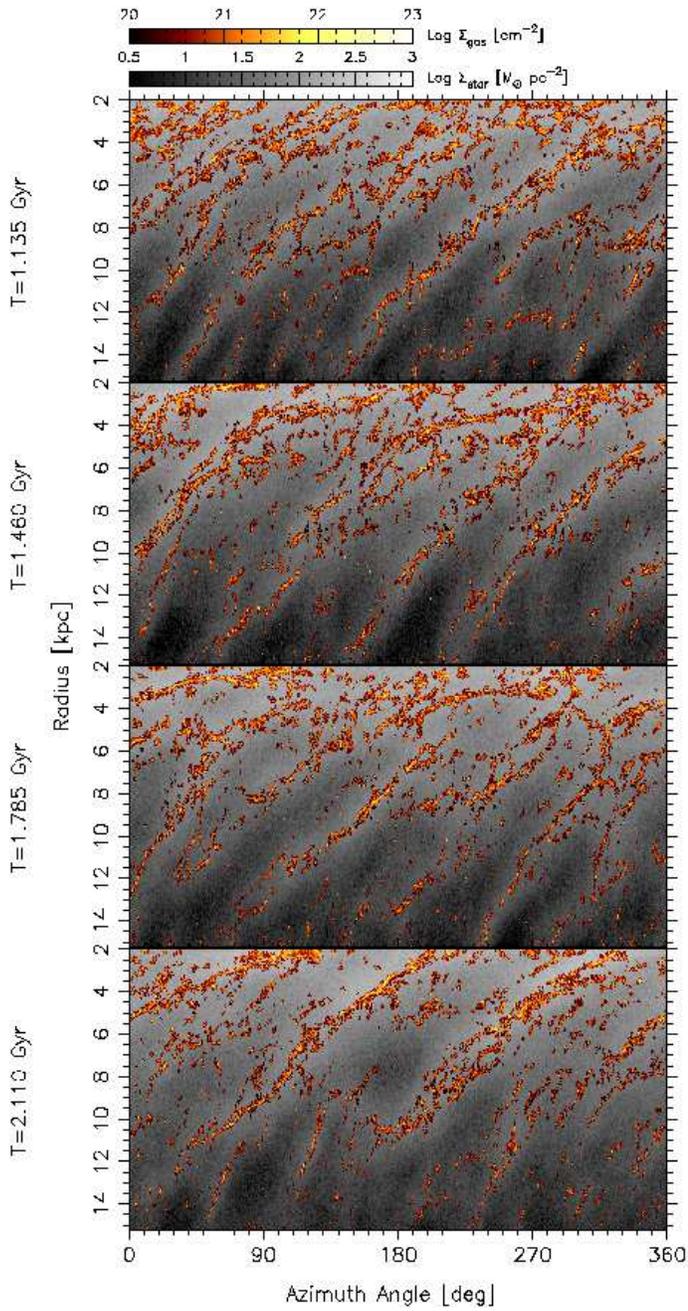}
 \caption{Cold gas ($T<100$ K), whose surface density is represented by the color,
 is overlaid on a stellar density shown by a gray-scale.
% {KW:  Need to be replaced} Distribution of cold gas ($T < 100 $ K, blue-green-red-purple) 
% and stars (yellow-brown) at $t=1.63 $ Myrs and $t= 1.83$ Myr. 
See also the movie in {\it the supplementary data.}}
   \label{fig3}
\end{center}
\end{figure}

%%% Figure 6 %%%%%%
\begin{figure*}[t]
\begin{center}
  \includegraphics[width=1.00\textwidth]{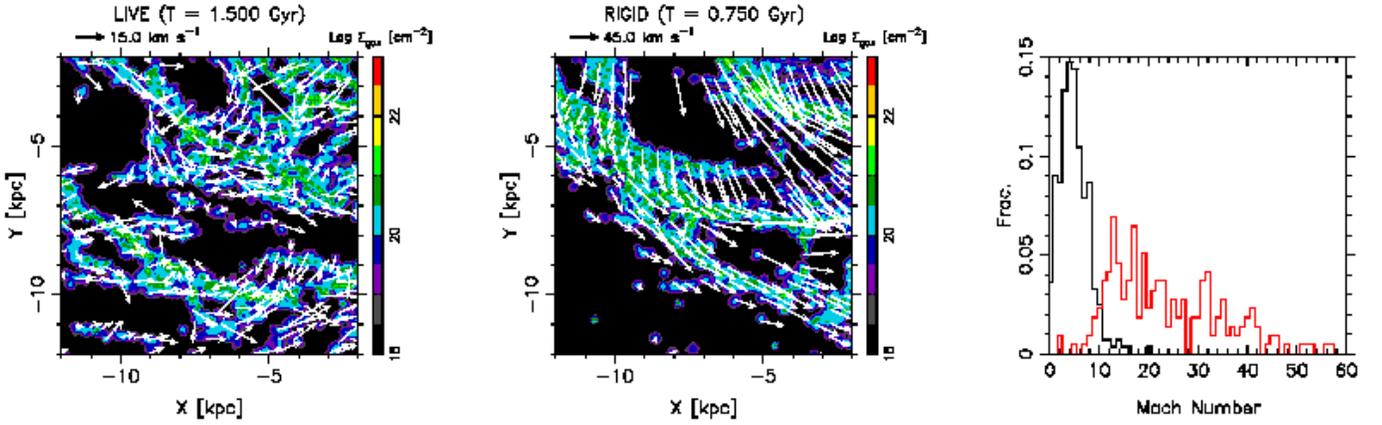}
 \caption{(left) Kinematics and surface density of the cold gas ($T < 100 $ K) in the live spiral model.
 Arrows represent velocities of the gas relative to the galactic rotation, 
 therefore they show approximately relative velocities to the local stellar spiral arms (see Fig. \ref{fig2}).
 Color represents gas density.
(middle) Same as the left panel, but for the rigid spiral model.  The velocities are 
on a rotating frame of the spiral potential. 
Note that 
unit length of the vectors in the two panels differ, and 
the velocities are much faster in the 
right two models than those in the live  model.  
(right) Frequency distributions of the Mach number  {of SPH particles $\cal{M}$ $ \equiv |v|/c_{\rm s}$ (see the text for the definition)
} in
the live spiral model (black) and rigid spiral model (red).}
%See also the movie in the supplementary data.}
   \label{fig6}
\end{center}
\end{figure*}

%%% Figure 7 %%%%%%
\begin{figure*}[t]
\begin{center}
  \includegraphics[width=0.65\textwidth]{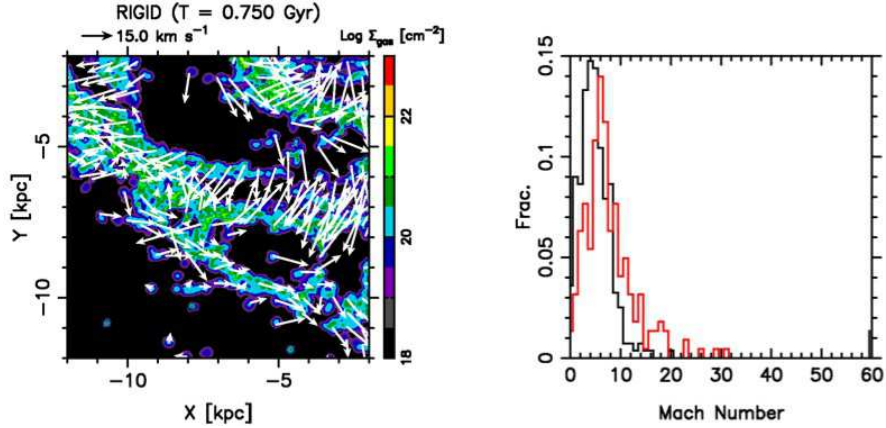}
 \caption{ Same as the middle and right panels of Fig. \ref{fig6}, but
 the velocities are defined on 
 the local galactic rotation.  }

   \label{fig7new}
\end{center}
\end{figure*}

%%% Figure 8   %%%%%%

\begin{figure*}[t]
\begin{center}
  \includegraphics[width=0.90\textwidth]{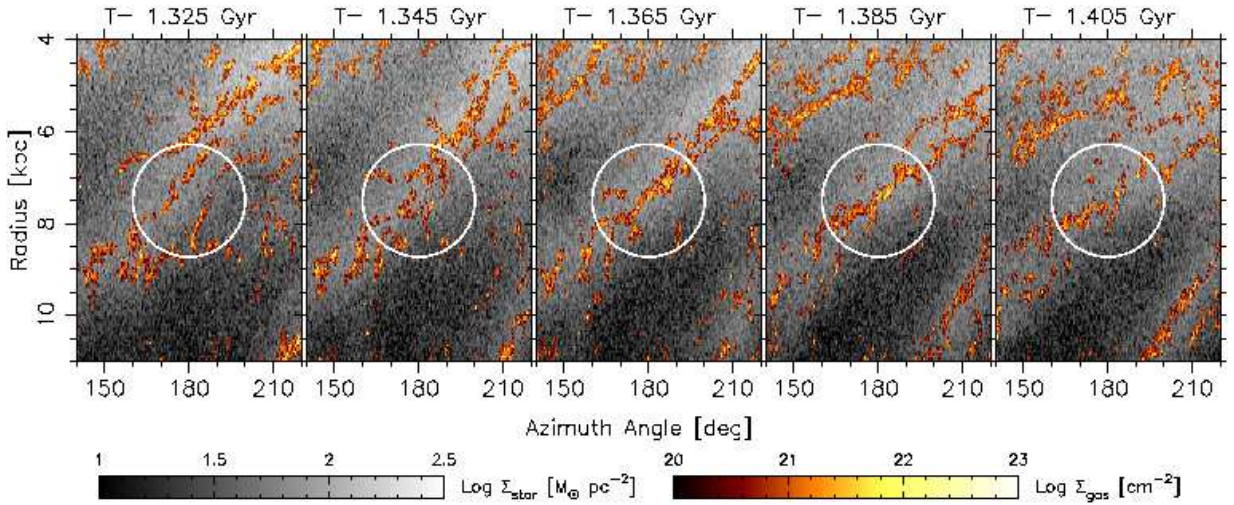}  
 \caption{Same as Fig. \ref{fig3}, but a close-up of a collision of gas clouds near the stellar arm every 20 Myrs. }
   \label{fig7}
\end{center}
\end{figure*}

%%% Figure 9 %%%%%%
\begin{figure*}[t]
% \vspace*{-2.0 cm}
\begin{center}
\includegraphics[width=0.90\textwidth]{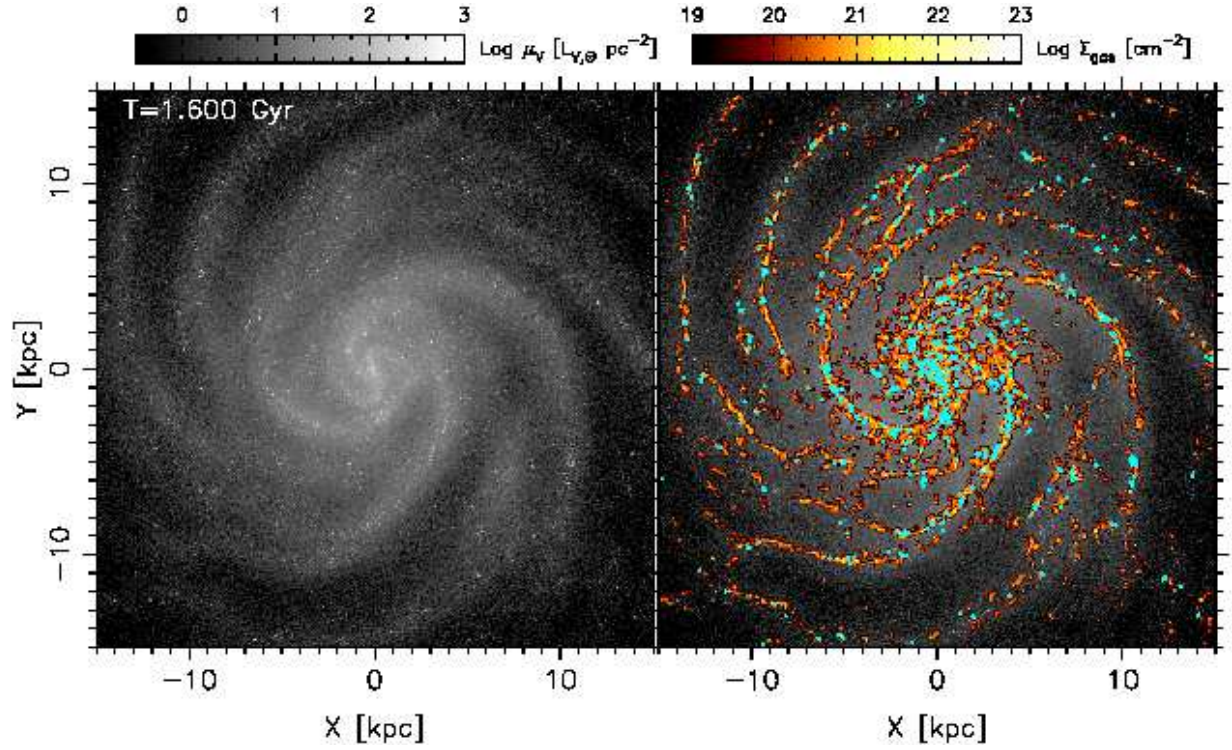}
% \vspace*{-1.0 cm}
 \caption{(Left) $V$-band surface luminosity calculated from stellar density with
 a population synthesis model. (Right) {Cold gas ($T<100$ K)} density and 
 star particles,  whose ages are  $< 30$ Myrs represented by light blue color,  formed from cold, dense gas are overlaid on 
 the left panel. }
   \label{fig8}
\end{center}
\end{figure*}

%%% Figure 10  %%%%%%
\begin{figure*}[t]
\begin{center}
  \includegraphics[width=0.90\textwidth]{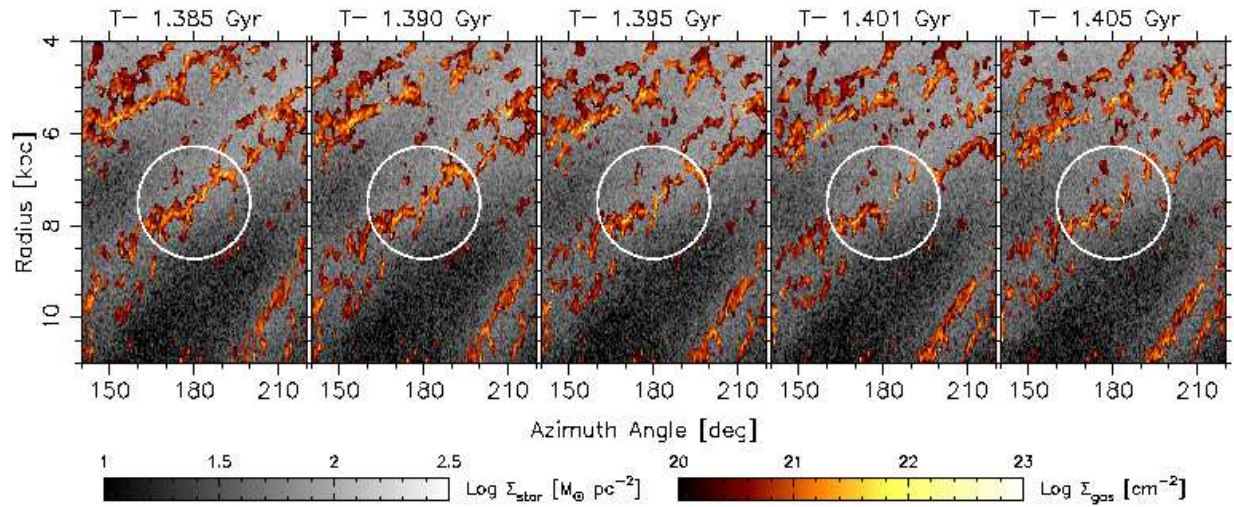}  
 \caption{Same as Fig. \ref{fig7}, but snapshot every 5 Myrs after $t = 1.385$ Gyrs. }
   \label{fig9a}
\end{center}
\end{figure*}

%%% Figure 11  %%%%%%
\begin{figure*}[t]
\begin{center}
  \includegraphics[width=0.90\textwidth]{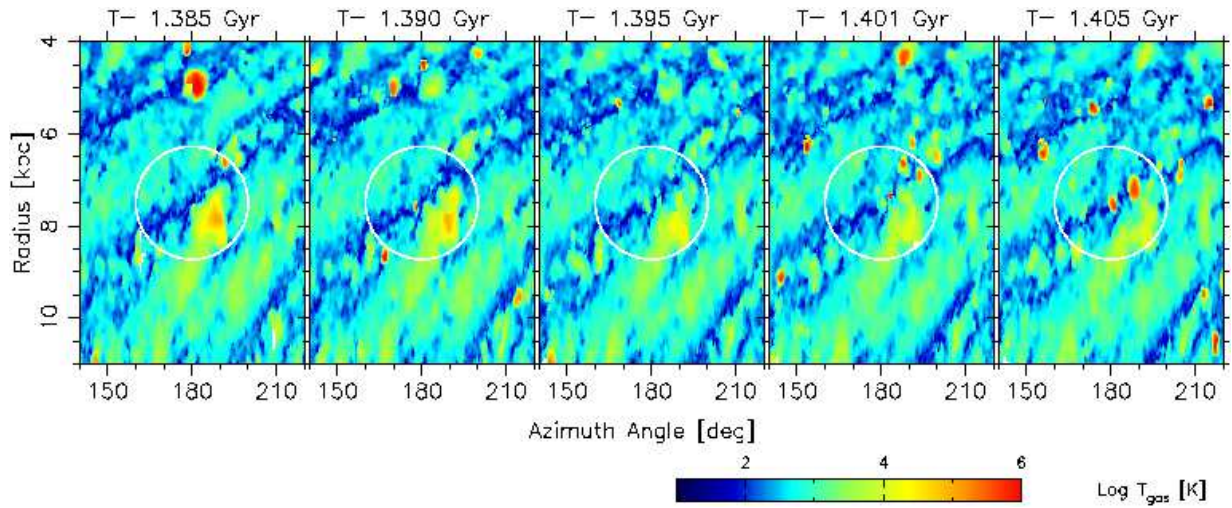}  
 \caption{Same as Fig. \ref{fig9a}, but temperature distribution.  The hot regions (red) are generated by supernova explosions.
 (see also the movie in {\it the supplementary data})}
   \label{fig9b}
\end{center}
\end{figure*}

%%% Figure 12  %%%%%%
\begin{figure*}[t]
\begin{center}
  \includegraphics[width=0.40\textwidth]{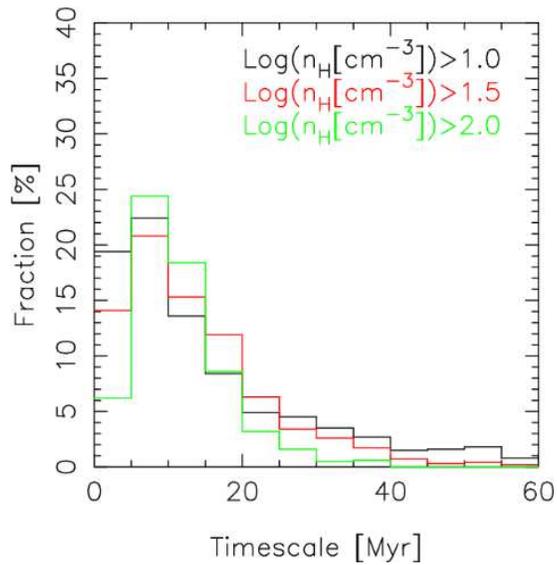}  
 \caption{Frequency distribution of `life time'  to keep number densities, $n_{\rm H} \ge 10, 10^{1.5}$ and $10^2$ cm$^{-3}$. }
   \label{fig11}
\end{center}
\end{figure*}

%\bibliographystyle{apj}   %
%\bibliography{ms}     %

\begin{thebibliography}{62}
\expandafter\ifx\csname natexlab\endcsname\relax\def\natexlab#1{#1}\fi

\bibitem[{{Baba} {et~al.}(2009){Baba}, {Asaki}, {Makino}, {Miyoshi}, {Saitoh},
  \& {Wada}}]{Baba+2009}
{Baba}, J., {Asaki}, Y., {Makino}, J., {Miyoshi}, M., {Saitoh}, T.~R., \&
  {Wada}, K. 2009, \apj, 706, 471

\bibitem[{{Baba} {et~al.}(2010){Baba}, {Saitoh}, \& {Wada}}]{Baba+2010}
{Baba}, J., {Saitoh}, T.~R., \& {Wada}, K. 2010, arXiv 1009.3096

\bibitem[{{Bertin} \& {Lin}(1996)}]{BertinLin1996}
{Bertin}, G., \& {Lin}, C.~C. 1996, {Spiral structure in galaxies a density
  wave theory}, ed. G.~{Bertin} \& C.~C. {Lin}

\bibitem[{{Binney} \& {Tremaine}(2008)}]{BinneyTremaine2008}
{Binney}, J., \& {Tremaine}, S. 2008, {Galactic Dynamics: Second Edition}, ed.
  {Binney, J.~\& Tremaine, S.} (Princeton University Press)

\bibitem[{{Dobbs} \& {Bonnell}(2006)}]{DobbsBonnell2006}
{Dobbs}, C.~L., \& {Bonnell}, I.~A. 2006, \mnras, 367, 873

\bibitem[{{Dobbs} {et~al.}(2010){Dobbs}, {Theis}, {Pringle}, \&
  {Bate}}]{Dobbs+2010}
{Dobbs}, C.~L., {Theis}, C., {Pringle}, J.~E., \& {Bate}, M.~R. 2010, \mnras,
  403, 625

\bibitem[{{Elmegreen}(1980)}]{Elmegreen1980}
{Elmegreen}, D.~M. 1980, \apj, 242, 528

\bibitem[{{Fioc} \& {Rocca-Volmerange}(1999)}]{FiocRocca1999}
{Fioc}, M., \& {Rocca-Volmerange}, B. 1999, arXiv:astro-ph/9912179

\bibitem[{{Fuchs} {et~al.}(2005){Fuchs}, {Dettbarn}, \&
  {Tsuchiya}}]{Fuchs+2005}
{Fuchs}, B., {Dettbarn}, C., \& {Tsuchiya}, T. 2005, \aap, 444, 1

\bibitem[{{Fujii} {et~al.}(2011){Fujii}, {Baba}, {Saitoh}, {Makino}, {Kokubo},
  \& {Wada}}]{Fujii+2011}
{Fujii}, M.~S., {Baba}, J., {Saitoh}, T.~R., {Makino}, J., {Kokubo}, E., \&
  {Wada}, K. 2011, \apj, 730, 109

\bibitem[{{Fujimoto}(1968)}]{Fujimoto1968}
{Fujimoto}, M. 1968, in IAU Symposium, Vol.~29, IAU Symposium, 453--+

\bibitem[{{Gerritsen} \& {Icke}(1997)}]{GerritsenIcke1997}
{Gerritsen}, J.~P.~E., \& {Icke}, V. 1997, \aap, 325, 972

\bibitem[{{Gingold} \& {Monaghan}(1977)}]{GingoldMonaghan1977}
{Gingold}, R.~A., \& {Monaghan}, J.~J. 1977, \mnras, 181, 375

\bibitem[{{Gratier} {et~al.}(2010){Gratier}, {Braine}, {Rodriguez-Fernandez},
  {Schuster}, {Kramer}, {Xilouris}, {Tabatabaei}, {Henkel}, {Corbelli},
  {Israel}, {van der Werf}, {Calzetti}, {Garcia-Burillo}, {Sievers}, {Combes},
  {Wiklind}, {Brouillet}, {Herpin}, {Bontemps}, {Aalto}, {Koribalski}, {van der
  Tak}, {Wiedner}, {R{\"o}llig}, \& {Mookerjea}}]{Gratier+2010}
{Gratier}, P. {et~al.} 2010, \aap, 522, A3+

\bibitem[{{Hernquist}(1993)}]{Hernquist1993}
{Hernquist}, L. 1993, \apjs, 86, 389

\bibitem[{{Inoue} \& {Inutsuka}(2009)}]{InoueInutsuka2009}
{Inoue}, T., \& {Inutsuka}, S. 2009, \apj, 704, 161

\bibitem[{{Jog} \& {Solomon}(1984)}]{JogSolomon1984}
{Jog}, C.~J., \& {Solomon}, P.~M. 1984, \apj, 276, 114

\bibitem[{{Kalnajs}(1972)}]{Kalnajs1972}
{Kalnajs}, A.~J. 1972, \aplett, 11, 41

\bibitem[{{Kalnajs}(1973)}]{Kalnajs1973}
---. 1973, Proceedings of the Astronomical Society of Australia, 2, 174

\bibitem[{{Kennicutt}(1998)}]{Kennicutt1998}
{Kennicutt}, Jr., R.~C. 1998, \apj, 498, 541

\bibitem[{{Kim} {et~al.}(2006){Kim}, {Kim}, \& {Ostriker}}]{Kim+2006}
{Kim}, C., {Kim}, W., \& {Ostriker}, E.~C. 2006, \apjl, 649, L13

\bibitem[{{Kim} {et~al.}(2010){Kim}, {Kim}, \& {Ostriker}}]{Kim+2010}
---. 2010, \apj, 720, 1454

\bibitem[{{Kim} \& {Ostriker}(2002)}]{KimOstriker2002}
{Kim}, W., \& {Ostriker}, E.~C. 2002, \apj, 570, 132

\bibitem[{{Kim} \& {Ostriker}(2006)}]{KimOstriker2006}
---. 2006, \apj, 646, 213

\bibitem[{{La Vigne} {et~al.}(2006){La Vigne}, {Vogel}, \&
  {Ostriker}}]{LaVigne+2006}
{La Vigne}, M.~A., {Vogel}, S.~N., \& {Ostriker}, E.~C. 2006, \apj, 650, 818

\bibitem[{{Lin} \& {Bertin}(1985)}]{LinBertin1985}
{Lin}, C.~C., \& {Bertin}, G. 1985, in IAU Symposium, Vol. 106, The Milky Way
  Galaxy, ed. {H.~van Woerden, R.~J.~Allen, \& W.~B.~Burton}, 513--530

\bibitem[{{Lin} \& {Shu}(1964)}]{LinShu1964}
{Lin}, C.~C., \& {Shu}, F.~H. 1964, \apj, 140, 646

\bibitem[{{Lucy}(1977)}]{Lucy1977}
{Lucy}, L.~B. 1977, \aj, 82, 1013

\bibitem[{{Martin} \& {Kennicutt}(2001)}]{MartinKennicutt2001}
{Martin}, C.~L., \& {Kennicutt}, Jr., R.~C. 2001, \apj, 555, 301

\bibitem[{{McMillan} \& {Dehnen}(2007)}]{McMillanDehnen2007}
{McMillan}, P.~J., \& {Dehnen}, W. 2007, \mnras, 378, 541

\bibitem[{{Navarro} {et~al.}(1997){Navarro}, {Frenk}, \&
  {White}}]{Navarro+1997}
{Navarro}, J.~F., {Frenk}, C.~S., \& {White}, S.~D.~M. 1997, \apj, 490, 493

\bibitem[{{Norman}(1978)}]{Norman1978}
{Norman}, C.~A. 1978, \mnras, 182, 457

\bibitem[{{Oh} {et~al.}(2008){Oh}, {Kim}, {Lee}, \& {Kim}}]{Oh+2008}
{Oh}, S.~H., {Kim}, W., {Lee}, H.~M., \& {Kim}, J. 2008, \apj, 683, 94

\bibitem[{{Onodera} {et~al.}(2010){Onodera}, {Kuno}, {Tosaki}, {Kohno},
  {Nakanishi}, {Sawada}, {Muraoka}, {Komugi}, {Miura}, {Kaneko}, {Hirota}, \&
  {Kawabe}}]{Onodera+2010}
{Onodera}, S. {et~al.} 2010, \apjl, 722, L127

\bibitem[{{Reid} {et~al.}(2009){Reid}, {Menten}, {Zheng}, {Brunthaler},
  {Moscadelli}, {Xu}, {Zhang}, {Sato}, {Honma}, {Hirota}, {Hachisuka}, {Choi},
  {Moellenbrock}, \& {Bartkiewicz}}]{Reid+2009b}
{Reid}, M.~J. {et~al.} 2009, \apj, 700, 137

\bibitem[{{Roberts}(1969)}]{Roberts1969}
{Roberts}, W.~W. 1969, \apj, 158, 123

\bibitem[{{Roberts} \& {Shu}(1972)}]{RobertsShu1972}
{Roberts}, Jr., W.~W., \& {Shu}, F.~H. 1972, \aplett, 12, 49

\bibitem[{{Robertson} \& {Kravtsov}(2008)}]{RobertsonKravtsov2008}
{Robertson}, B.~E., \& {Kravtsov}, A.~V. 2008, \apj, 680, 1083

\bibitem[{{Saitoh} {et~al.}(2008){Saitoh}, {Daisaka}, {Kokubo}, {Makino},
  {Okamoto}, {Tomisaka}, {Wada}, \& {Yoshida}}]{Saitoh+2008}
{Saitoh}, T.~R., {Daisaka}, H., {Kokubo}, E., {Makino}, J., {Okamoto}, T.,
  {Tomisaka}, K., {Wada}, K., \& {Yoshida}, N. 2008, \pasj, 60, 667

\bibitem[{{Saitoh} {et~al.}(2009){Saitoh}, {Daisaka}, {Kokubo}, {Makino},
  {Okamoto}, {Tomisaka}, {Wada}, \& {Yoshida}}]{Saitoh+2009}
---. 2009, \pasj, 61, 481

\bibitem[{{Saitoh} \& {Makino}(2010)}]{SaitohMakino2010}
{Saitoh}, T.~R., \& {Makino}, J. 2010, \pasj, 62, 301

\bibitem[{{Salpeter}(1955)}]{Salpeter1955}
{Salpeter}, E.~E. 1955, \apj, 121, 161

\bibitem[{{Scoville} {et~al.}(2001){Scoville}, {Polletta}, {Ewald}, {Stolovy},
  {Thompson}, \& {Rieke}}]{Scoville+2001}
{Scoville}, N.~Z., {Polletta}, M., {Ewald}, S., {Stolovy}, S.~R., {Thompson},
  R., \& {Rieke}, M. 2001, \aj, 122, 3017

\bibitem[{{Sellwood}(2000)}]{Sellwood2000}
{Sellwood}, J.~A. 2000, \apss, 272, 31

\bibitem[{{Sellwood}(2010)}]{Sellwood2010}
---. 2010, ArXiv e-prints: 1006.4855

\bibitem[{{Sellwood}(2011)}]{Sellwood2011}
---. 2011, \mnras, 410, 1637

\bibitem[{{Sellwood} \& {Binney}(2002)}]{SellwoodBinney2002}
{Sellwood}, J.~A., \& {Binney}, J.~J. 2002, \mnras, 336, 785

\bibitem[{{Sellwood} \& {Carlberg}(1984)}]{SellwoodCarlberg1984}
{Sellwood}, J.~A., \& {Carlberg}, R.~G. 1984, \apj, 282, 61

\bibitem[{{Sellwood} \& {Sparke}(1988)}]{SellwoodSparke1988}
{Sellwood}, J.~A., \& {Sparke}, L.~S. 1988, \mnras, 231, 25P

\bibitem[{{Shetty} \& {Ostriker}(2006)}]{ShettyOstriker2006}
{Shetty}, R., \& {Ostriker}, E.~C. 2006, \apj, 647, 997

\bibitem[{{Shu} {et~al.}(1972){Shu}, {Milione}, {Gebel}, {Yuan}, {Goldsmith},
  \& {Roberts}}]{Shu+1972}
{Shu}, F.~H., {Milione}, V., {Gebel}, W., {Yuan}, C., {Goldsmith}, D.~W., \&
  {Roberts}, W.~W. 1972, \apj, 173, 557

\bibitem[{{Shu} {et~al.}(1973){Shu}, {Milione}, \& {Roberts}}]{Shu+1973}
{Shu}, F.~H., {Milione}, V., \& {Roberts}, Jr., W.~W. 1973, \apj, 183, 819

\bibitem[{{Spaans} \& {Norman}(1997)}]{SpaansNorman1997}
{Spaans}, M., \& {Norman}, C.~A. 1997, \apj, 483, 87

\bibitem[{{Springel}(2010)}]{Springel2010}
{Springel}, V. 2010, \araa, 48, 391

\bibitem[{{Struck} {et~al.}(2011){Struck}, {Dobbs}, \& {Hwang}}]{Struck+2011}
{Struck}, C., {Dobbs}, C.~L., \& {Hwang}, J. 2011, ArXiv e-prints

\bibitem[{{Sugimoto} {et~al.}(1990){Sugimoto}, {Chikada}, {Makino}, {Ito},
  {Ebisuzaki}, \& {Umemura}}]{Sugimoto+1990}
{Sugimoto}, D., {Chikada}, Y., {Makino}, J., {Ito}, T., {Ebisuzaki}, T., \&
  {Umemura}, M. 1990, \nat, 345, 33

\bibitem[{{Toomre}(1981)}]{Toomre1981}
{Toomre}, A. 1981, in Structure and Evolution of Normal Galaxies, ed. S.~M.
  {Fall} \& D.~{Lynden-Bell}, 111--136

\bibitem[{{Vazquez-Semadeni} {et~al.}(2011){Vazquez-Semadeni}, {Banerjee},
  {Gomez}, {Hennebelle}, {Duffin}, \& {Klessen}}]{Vazquez-Semadeni2011}
{Vazquez-Semadeni}, E., {Banerjee}, R., {Gomez}, G., {Hennebelle}, P.,
  {Duffin}, D., \& {Klessen}, R.~S. 2011, ArXiv e-prints: 1101.3384

\bibitem[{{Wada}(2008)}]{Wada2008}
{Wada}, K. 2008, \apj, 675, 188

\bibitem[{{Wada} \& {Koda}(2004)}]{WadaKoda2004}
{Wada}, K., \& {Koda}, J. 2004, \mnras, 349, 270

\bibitem[{{Wada} \& {Norman}(2001)}]{WadaNorman2001}
{Wada}, K., \& {Norman}, C.~A. 2001, \apj, 547, 172

\bibitem[{{Woodward}(1975)}]{Woodward1975}
{Woodward}, P.~R. 1975, \apj, 195, 61

\end{thebibliography}

\end{document}